\documentclass[sigconf]{acmart}

\usepackage{enumitem}
\usepackage{soul}
\usepackage{color}
\usepackage{microtype}
\usepackage{subcaption}

\copyrightyear{2023}
\acmYear{2023}
\setcopyright{rightsretained}
\acmConference[IUI '23]{28th International Conference on Intelligent User Interfaces}{March 27--31, 2023}{Sydney, NSW, Australia}
\acmBooktitle{28th International Conference on Intelligent User Interfaces (IUI '23), March 27--31, 2023, Sydney, NSW, Australia}
\acmDOI{10.1145/3581641.3584034}
\acmISBN{979-8-4007-0106-1/23/03}

\begin{document}

\title{\scim{}: Intelligent Skimming Support for Scientific Papers}

\author{Raymond Fok}
\email{rayfok@cs.washington.edu}
\affiliation{
  \institution{University of Washington}
  \city{Seattle}
  \state{Washington}
  \country{USA}
}

\author{Hita Kambhamettu}
\email{hitakam@seas.upenn.edu}
\affiliation{
  \institution{University of Pennsylvania}
  \city{Philadelphia}
  \state{Pennsylvania}
  \country{USA}
}

\author{Luca Soldaini}
\email{lucas@allenai.org}
\affiliation{
  \institution{Allen Institute for AI}
  \city{Seattle}
  \state{Washington}
  \country{USA}
}

\author{Jonathan Bragg}
\email{jbragg@allenai.org}
\affiliation{
  \institution{Allen Institute for AI}
  \city{Seattle}
  \state{Washington}
  \country{USA}
}

\author{Kyle Lo}
\email{kylel@allenai.org}
\affiliation{
  \institution{Allen Institute for AI}
  \city{Seattle}
  \state{Washington}
  \country{USA}
}

\author{Marti A. Hearst}
\email{hearst@berkeley.edu}
\affiliation{
  \institution{UC Berkeley}
  \city{Berkeley}
  \state{California}
  \country{USA}
}

\author{Andrew Head}
\email{head@seas.upenn.edu}
\affiliation{
  \institution{University of Pennsylvania}
  \city{Philadelphia}
  \state{Pennsylvania}
  \country{USA}
}

\author{Daniel S. Weld}
\email{danw@allenai.org}
\affiliation{
  \institution{Allen Institute for AI \&\ \\ University of Washington}
  \city{Seattle}
  \state{Washington}
  \country{USA}
}

\renewcommand{\shortauthors}{Fok et al.}

\begin{abstract}
Scholars need to keep up with an exponentially increasing flood of scientific papers. To aid this challenge, we introduce \scim{}, a novel intelligent interface that helps experienced researchers skim -- or rapidly review -- a paper to attain a cursory understanding of its contents. \scim{} supports the skimming process by highlighting salient paper contents in order to direct a reader's attention. The system's highlights are faceted by content type, evenly distributed across a paper, and have a density configurable by readers at both the global and local level. We evaluate \scim{} with both an in-lab usability study and a longitudinal diary study, revealing how its highlights facilitate the more efficient construction of a conceptualization of a paper. We conclude by discussing design considerations and tensions for the design of future intelligent skimming tools.
\end{abstract}

\newcommand{\needcite}[1]{[XXX #1 XXX]}
\renewcommand{\comment}[1]{}

\newcommand\dan[1]{\textcolor{red}{#1}}
\newcommand\andrew[1]{\textcolor{brown}{#1}}
\newcommand\ray[1]{\textcolor{green}{#1}}
\newcommand\marti[1]{\textcolor{teal}{#1}}
\newcommand{\revise}[1]{\textcolor{blue}{#1}}

\newcommand{\bug}
    {\mbox{\rule{2mm}{2mm}}}
\newcommand{\DSW}[1]
    {\bug \footnote{\textcolor{red}{\textit{DSW: #1}}}}

\newcommand{\scim}{\textsc{Scim}}
\definecolor{lightblue}{rgb}{.90,.95,1}
\DeclareRobustCommand{\hlmethod}[1]{{\sethlcolor{lightblue}\hl{#1}}}

\definecolor{methodblue}{RGB}{182,219,240}
\definecolor{resultred}{RGB}{247,187,204}
\definecolor{noveltyorange}{RGB}{249,200,181}
\definecolor{objectivegreen}{RGB}{174,255,167}



\maketitle

\section{Introduction}


\begin{figure*}[t]
        \centering 
        \includegraphics[width=0.98\textwidth]{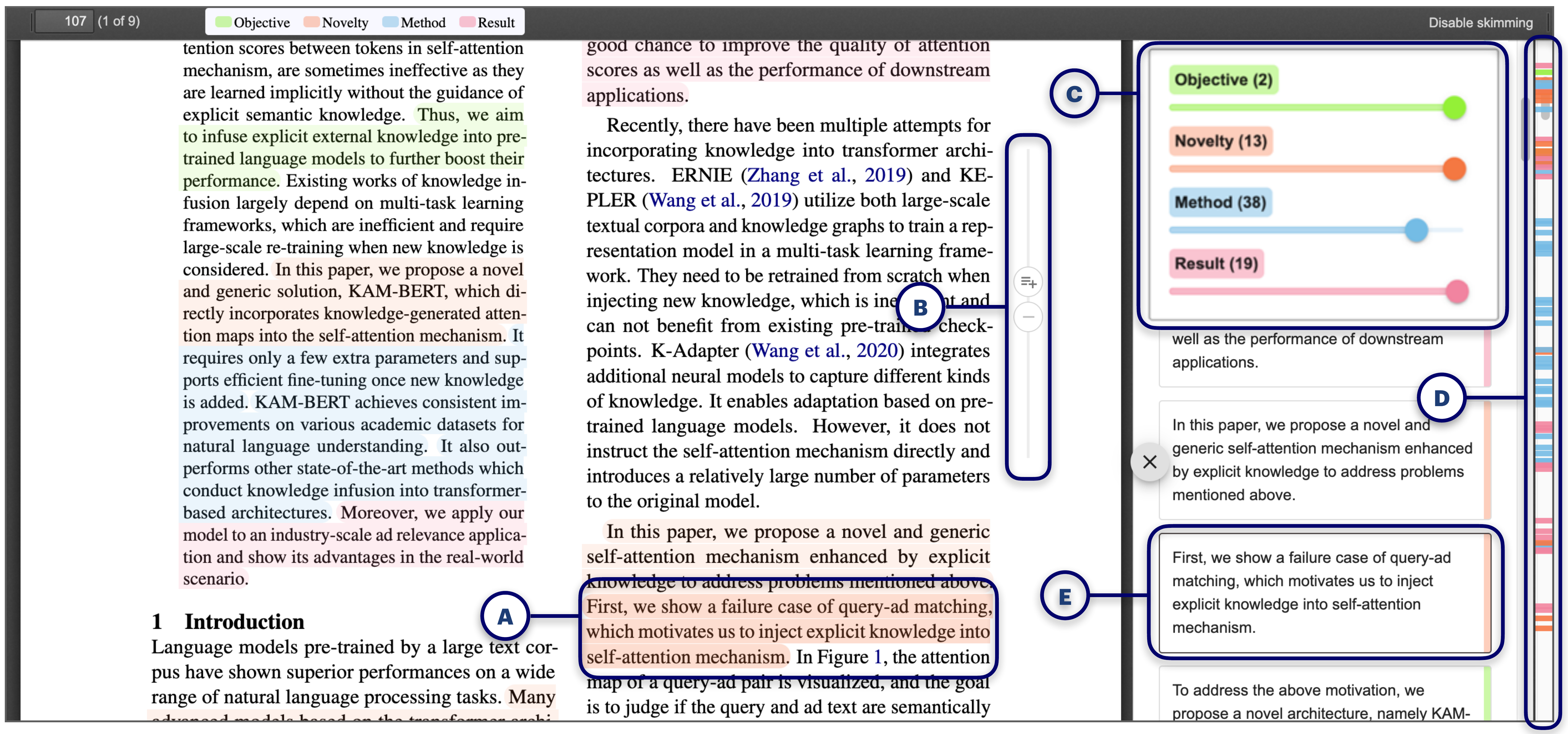}
        \caption{
        \scim{} is an intelligent reading interface for skimming scientific papers. To help readers develop a broad overview of content in a paper, \scim{} intelligently highlights passages (A). The passages are colorized to indicate the rhetorical role of the passage, i.e., whether it describes the research's objectives, novelty, methods, and results. Highlights are distributed throughout the text to support a holistic skim. Readers can request additional (or fewer) highlights by using paragraph-local (B) and paper-wide (C) controls. To understand where to find information of a certain kind, readers can glance at highlight markers in the scroll bar (D). Readers can also collect an overview of the paper by reviewing highlighted passages in a sidebar (E).}
        \Description{The Scim interface, with individual features annotated. Further details of the interface is described in the paper text.}
        \label{fig:user_interface}
    \end{figure*}

With the rise of knowledge work and a contemporaneous explosion of information, experts are expected to sift through and make sense of large volumes of rapidly evolving information. One domain where this trend is particularly pronounced is scientific research. Researchers spend a tremendous amount of effort staying up to date with the literature. They do so by regularly undertaking the tasks of foraging for papers, skimming or reading those deemed most relevant, and integrating knowledge gained from reading into their personal records.

Skimming is a critical task, and requires researchers to quickly review the contents of a paper to develop a cursory understanding of its contents. While faster than reading, skimming achieves a coarser view of papers' contents. With the shift of scientific publishing from paper to digital online publications, the practice of skimming has become yet more widespread~\cite{liu_reading_2005, tenopir_electronic_2009}. Despite the pervasiveness of skimming as a practice for reviewing papers \cite{rayner_so_2016}, skimming is not easy~\cite{maxwell_skimming_1972, duggan_text_2009}. Skimming may devolve into reading should a reader find themselves drawn into the details of a passage. Even for experienced readers, skimming requires attention to make strategic choices of what to read, where, and when to stop reading. 

In this paper, we investigate how an intelligent user interface can help both novice and expert researchers skim scientific papers more efficiently. Today, techniques from Artificial Intelligence are increasingly used in search tools over the scholarly literature (cf.~\cite{beel_google_2009,ammar_construction_2018}) and in scientific reading applications (e.g.,~\cite{head_augmenting_2021, august_paperplain_2022}). This paper explores how intelligent tools can facilitate the task of skimming, the seam between searching and reading.

As a starting point, we ask how judicious use of automatic highlighting can be presented in a tool to help readers direct their attention while they skim. To gain inspiration for designing such a tool, we conducted formative studies with researchers, including interviews, observations, and pilot studies of prototype highlighting tools. In these studies, we found that readers desired highlights that  cover diverse content, are evenly distributed across a paper, and  capture important paper content. These studies reveal a tension between reader expectations and system design, because it is not always possible to highlight according to passage importance while achieving a desirable distribution of highlights. Readers also desired some influence over the quantity and distribution of highlights within a paper.

We incorporated these insights into the design of \scim{},\footnote{Code available at \url{https://github.com/rayfok/scim}} an intelligent user interface for skimming scientific papers (Figure~\ref{fig:user_interface}). To address readers' needs around highlighting, \scim{} highlights passages in the following ways. First, passages are highlighted with distinct colors for each of four diverse kinds of content sought by readers: research objectives, novel aspects of the research, methodology, and results. Second, \scim{} aims to support an evenly-distributed skim of a paper, highlighting passages in a way that most paragraphs contain at least one highlighted sentence. Finally, \scim{} lets readers customize the number of highlights in a paper, both across an entire paper and within individual paragraphs.

We conducted two studies to evaluate \scim{}'s utility as an intelligent skimming tool. First, we performed a lab usability study to understand how \scim{} affects readers' ability to search for specific kinds of information in a paper. When using \scim{}, readers located the desired information in significantly less time compared to a standard document reader, with comparable effort and accuracy. Second, to understand more realistic usage, we conducted a two-week-long diary study. In this study, readers found \scim{} particularly useful when skimming text-dense passages with few visuals, or when skimming a paper that fell outside their area of expertise. \scim{} became more usable over time as readers became accustomed to the highlights. The study also suggests how skimming assistants could be improved in the future, for instance by highlighting passages that provide background for later highlighted passages, and integrating highlights with the typographical emphases authors may have already provided, such as boldface font and text formatting.

The \scim{} system and its accompanying studies offer a vision of how applications can help readers skim texts like scientific papers with intelligent highlighting. Altogether, this paper contributes:
\begin{itemize}
    \item Design goals to guide the design of intelligent, highlight-based skimming user interfaces, based on formative interviews and preliminary usability studies of a prototype tool.
    \item \scim{}, an intelligent skimming interface for scientific papers that highlights passages in a way that balances importance, diversity, and distribution of content, and affords control at both the paper and paragraph level.
    \item A reference implementation of \scim{}'s end-to-end paper analysis pipeline, including a language model for classifying salient sentences fine-tuned using a data programming approach, and post-processing heuristics for improving accuracy and achieving well-distributed highlights.
    \item Insights into the strengths and limitations of \scim{} based on a lab usability study and a longitudinal diary study.
\end{itemize}
\section{Related Work}
In this section, we first introduce motivating insights about the process of skimming, and then we review tools and techniques that have been introduced to support skimming.

\subsection{The Skimming Process}
In the literature, skimming is characterized as a form of rapid reading in which the goal is to get a general idea of the text or visual content, typically by focusing on information relevant to one's goals and skipping over irrelevant information~\cite{masson_conceptual_1983, rayner_so_2016}. Skimming is a necessary and useful skill for researchers. As the number of published papers increases year over year and papers have moved from print into digital media, scholars have tended toward reading more papers and spending less time on each, likely doing so by skimming~\cite{tenopir_electronic_2009, liu_reading_2005}. 

The psychology literature describes skimming as a cognitively demanding task. In this task, readers incrementally build a mental model of the text and integrate information across sentences as they read~\cite{rayner_so_2016, rapp_dynamic_2005, tashman_active_2011}. Generally, readers are not accurate at identifying goal-relevant information within text. Skimming is also physically demanding---limitations in the oculomotor system, which is responsible for controlling eye movements, preclude rapid, accurate placements of eye gaze for extended periods of time, such as when a reader skims a long document~\cite{masson_cognitive_1982, masson_conceptual_1983}.

Amidst the challenges of skimming, success is often determined by a reader's ability to ``satisfice''~\cite{reader_allocating_2007, duggan_text_2009, duggan_skim_2011}. Satisficing is a skim reading strategy where a reader sets a threshold of how useful information should be to deserve their attention, and if a unit of text falls below that threshold, they skip to the next unit of text. Studies have found that readers tend to spend more time at the beginning of paragraphs, the top of pages, and the beginning of documents~\cite{duggan_text_2009}, perhaps in part because this information is often believed to have high relevance.

One study of skimming for scientific document triage found readers were hasty and incomplete, with readers scrolling through documents quickly and paying attention to highly visual content and section headers~\cite{loizides_empirical_2009}. Scientific documents are laden with visual content, typographical cues (e.g., italicized, bold, or colored text), and structural information. Readers draw on document features to support rapid comprehension via these macro- and micro-structures~\cite{carrell_facilitating_1985, lacroix_macrostructure_1999, machulla_skimreading_2018} and visual content~\cite{yi_qndreview_2014, how_keshav_2007}. In this paper, we explore how automated assistance may support skimming by cueing readers towards significant sentences that might otherwise be missed. \scim's use of highlighting lets readers continue to pay attention to traditional visual and structural landmarks, while also heeding the passages highlighted by the skimming assistant.

\subsection{Tools for Reading and Skimming}
Researchers have long sought to equip readers with tools to support and augment their cognition while reading documents. The nascent days of human-computer interaction saw the introduction of augmented reading interfaces to support the reading process, including fluid documents that provided contextual access to supplemental information between lines of text~\cite{chang_negotiation_1998}, fluid hypertext~\cite{zellweger_fluid_1998}, visualizations for social annotations within papers~\cite{hill_edit_1992}, and affordances for annotating papers and jumping readers to passages of interest~\cite{graham_readers_1999, schilit_beyond_1999}. Since then, several approaches have been proposed to support the various aspects of reading, such as document navigation and comprehension.

\subsubsection{Modified Scrolling Interactions}
One line of research sought to facilitate the rapid exploration of long documents by modifying the behavior of reading interfaces during scrolling. Applications of content-aware scrolling were used to redefine the presentation order of content within a document~\cite{ishak_content-aware_2006}, provide pseudo-haptic feedback when scrolling past relevant information~\cite{kim_content-aware_2014}, and dynamically resize document headings within paper thumbnails in a document viewer~\cite{buchanan_improving_2008}. The Spotlights project implemented an attention allocation technique which pinned headings and figures as static overlays to a document as it was continuously scrolled~\cite{lee_spotlights_2016}.

\subsubsection{Typographical Cueing}
Another approach involved augmenting reading interfaces with typographical cues, e.g., highlighting. Studies in cognitive psychology have found visual cueing mechanisms can be effective in focusing reader attention~\cite{chi_visual_2007} and improving retention of material~\cite{fowler_effectiveness_1974, rickards_notetaking_1980}. The Semantize system used highlights to visualize sentiment within a document, and underlined words with positive or negative sentiment in different colors~\cite{wecker_semantize_2014}. The ScentHighlights system used highlights to identify conceptually relevant text based on a user's query~\cite{chi_scenthighlights_2005}. The HiText technique introduced dynamic graded highlighting of sentences within a document in accordance with their salience~\cite{yang_hitext_2017}. Modern reading interfaces also commonly support readers in marking regions of interest with a document with highlights or free-text annotations. The pervasiveness of highlighting as a technique for drawing readers' attention can be attributed to the von Restorff isolation effect, which states an item isolated against a homogeneous background will be more likely to be attended to and remembered~\cite{von_restorff_uber_1933, hunt_subtlety_1995}. Studies have since found evidence of this effect on the visual foraging behavior of readers, finding that highlights attract about half of the total number of fixations within a document, and    readers' eyes are often drawn to them~\cite{chi_visual_2007}.

\subsubsection{Document Augmentations}
Beyond typographical cues, other reading interface augmentations exist to specifically support the reading of scientific papers. For instance, online paper providers like Springer, PubMed, and Semantic Scholar provide readers with in-context citation information. Experimental systems have linked document text to marks within charts~\cite{kong_extracting_2014} and cells within tables~\cite{kim_facilitating_2018}, generated on-demand visualizations based on text within the paper~\cite{badam_elastic_2019}, augmented static visualizations with animated~\cite{grossman_your_2015} or interactive~\cite{masson_chameleon_2020} overlays, and provided in-context definitions for nonce words~\cite{head_augmenting_2021}. We design \scim{} with inspiration from many of these prior reading interfaces. Extending those prior systems that use highlighting, \scim{} not only automatically extracts salient sentences, but also classifies each highlight into common categories of information needs.

\subsubsection{Summarization}
An alternative method to skimming a full paper is to read a shortened representation of the paper's content in the form of a summary. An author-provided summary is de facto included with each paper as an abstract, which researchers often read before continuing to the rest of the paper. Automated summarization has garnered significant interest from the natural language processing community, and extractive and abstractive methods for generating summaries from long-form documents have been developed over the years~\cite{aggarwal_survey_2012, sefid_scibertsum_2022}. Some methods have even been proposed for generating extreme (single sentence) summaries, called \textsc{TLDR}s, from full papers~\cite{cachola_tldr_2020}.

However, summaries are often unsatisfactory. Despite recent advances, automated summaries remain error-prone, susceptible to hallucination~\cite{zhao_reducing_2020}, and unreliable as a standalone replacement for reading a paper itself. Furthermore, summaries do not provide readers with the ability to quickly interact with the full paper. As readers' goals and interests change while reading, they may wish to explore certain sections in further detail. Unlike summaries, augmented reading interfaces naturally retain the context of the paper. In this work, we present automatically extracted salient paper content as faceted highlights within a carefully-designed augmented reading interface to provide the interactivity and context lacking in standalone summaries.
\section{Design Goals} \label{sec:design}
To better understand how to design usable, intelligent skimming interfaces, we used an iterative design process that began with interviews and observations of academic researchers (referred to as \textit{readers}), and continued into an evaluation of an early prototype of \scim{}. We first describe that design process (Section~\ref{sec:formative_methods}), and then distill the lessons learned from this formative research into a set of design goals to guide the design intelligent, highlighting-based skimming support tools (Section~\ref{sec:design_motivations}). 

\subsection{Design Methodology} \label{sec:formative_methods}

\subsubsection{Formative interviews and observations}
We conducted formative study sessions with eight readers (F1--8) to better understand how they skim scientific papers. All readers belonged to the target user group for \scim{}, and were either graduate students or academic faculty. Readers were first observed as they skimmed a paper of their choice, and then asked to describe their skimming process, including goals they held while skimming, strategies they employed, and any aspects of skimming they found difficult or tedious.

\subsubsection{Prototype development and evaluation}
A prototype of \scim{} was iteratively designed and developed based on our formative interviews and observations. While many kinds of tools could support skimming, our design exploration focused specifically on skimming aids which incorporate intelligent highlights.

The prototype was similar to the version of the \scim{} system described in Section~\ref{sec:scim_ui}, with a few differences. First, the prototype's highlighting policy was different, resulting in fewer highlighted passages, and a less uniform distribution of highlights. Second, the prototype had no paragraph-level or facet-specific controls for the number of highlights, but rather only paper-level controls on the number of highlights and toggle switches for individual facets.

Two usability studies were conducted with this prototype. Thirteen readers (E1--13) were recruited from university mailing lists, and via direct outreach following purposive and snowball sampling approaches. Sessions in both studies were one hour in length and conducted on the Zoom platform. In both studies, readers skimmed papers with \scim{} for a limited amount of time and completed a task demonstrating their understanding of the paper, for instance outlining the paper or answering questions about the paper. Afterward, readers were asked to comment on their interactions with \scim{} and what aspects of the system required improvement.

\subsubsection{Synthesis}
One author conducted analyzed data from the formative study and preliminary evaluations following a thematic analysis methodology~\cite[Ch. 5]{blandford_qualitative_2016}. Notes and transcripts from study sessions were analyzed for themes and supporting evidence. Themes were validated through discussion and review with a second author. Those themes that provided actionable guidance for design are reported in the next section.

\subsection{Design Goals}\label{sec:design_motivations}
We introduce seven design goals for intelligent highlight-based skimming interfaces, based our formative research.

\emph{D1. Augment readers' skimming practices.}
Readers described myriad strategies they already used to skim papers. One common strategy was to first read the abstract and introduction of a paper. Then, readers consulted other key material in the paper, including bulleted lists of contributions (F1, F4, F6), summaries of results (F1--3), and conclusions (F1, F3, F6, F7). Readers also employed strategies particular to their goals, the paper, or their level of comfort with the paper. Readers relied on various visible cues in the text to help them identify important information, including typographical cues (e.g., italics, boldface) (F3, F6), structural cues (e.g., section headers) (F2, F6), visuals (e.g., figures and tables) (F1, F2, F4, F6, F8), and text position (e.g., inspecting the first sentences of paragraphs) (F2, F3, F6). We believe skimming interfaces should not impede or replace these reading strategies.

\emph{D2. Highlight diverse kinds of content.}
Readers' skimming goals were diverse. For instance, some readers sought to learn specific techniques introduced in a paper (F1), and others wished to understand a paper's relationship to prior research, or discover new research directions (F2--4, F7). Some desired a high-level understanding suitable for discussing the paper with colleagues (F3, F7). These goals influenced readers' skimming strategies, leading them to look for answers to different sorts of questions. We suggest skimming tools should support readers' diverse goals by enabling review of varied aspects of paper contents.

\emph{D3. Support skimming in the lengthy middle sections of the paper.}
Readers noted that while one recommended strategy for skimming is to read the beginning and ends of paragraphs, important content may reside in the middle of paragraphs. When asked to skim, we often observed readers transitioning into a deep read of some passages in the paper (F1, F3, F5). We propose that skimming tools should help readers identify important passages which conventional strategies do not reach, such as content in the middle of paragraphs and in the middle of the paper. 

\emph{D4. Minimize distraction.}
Without careful visual design, an augmented reading tool can occlude text or misdirect readers' attention. Our early prototypes incorporated a variety of text highlighting techniques, including underlines, lowlighting unimportant paper contents (inspired by ScholarPhi~\cite{head_augmenting_2021}), and highlighting text by setting its background color. Underlining was too subtle to consistently catch the reader's eye. Lowlighting tended to distract readers, requiring additional effort to read lowlighted content. Highlighting was chosen for its familiar use in documents, with the colors tuned to distinguish the categories of text and minimal contrast to avoid an unpleasant visual pop-out effect. We suggest that other designers similarly aim to minimize the visual distraction introduced by design interventions.


\begin{figure}[t]
        \centering
        \includegraphics[width=0.48\textwidth]{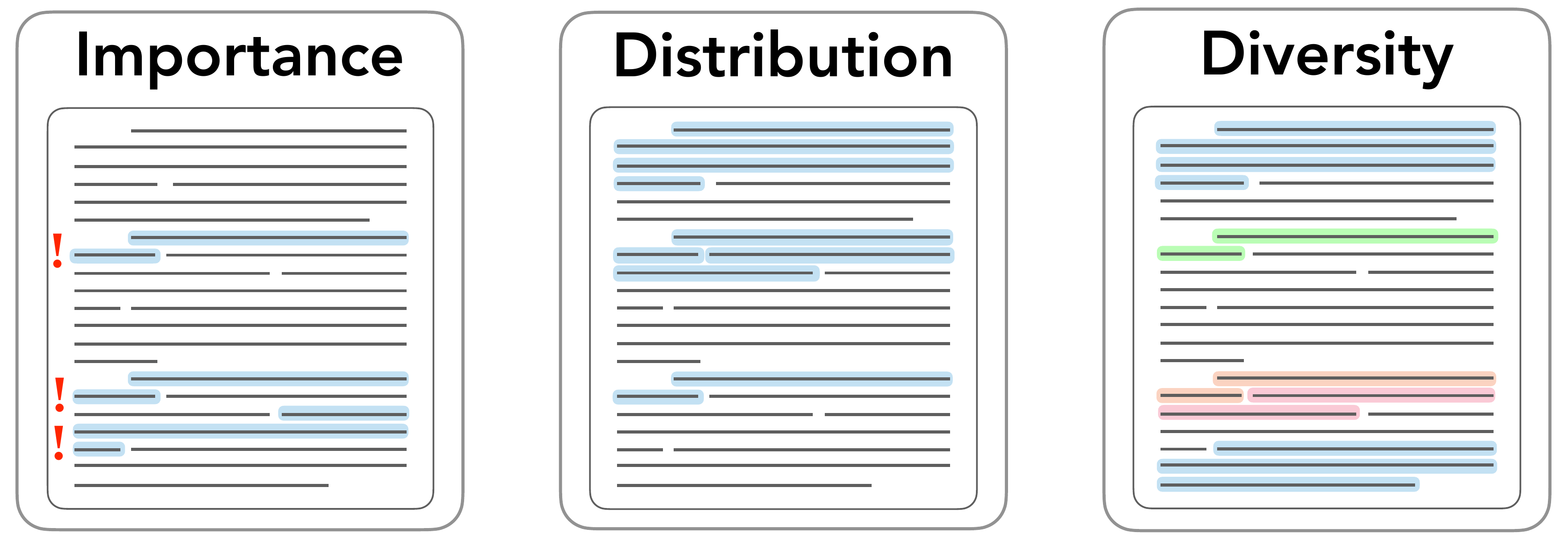}
        \caption{Our formative research revealed that intelligent highlights need to do more than pointing readers to important content. They should also be \textit{well-distributed} throughout a paper (D3, D5) and steer readers towards \textit{diverse} content types (D2).}
        \Description{Three paper icons displayed side by side. The icons are labeled importance, distribution, and diversity, left to right. The first paper icon is labeled importance, and shows three sentences highlighted, with an exclamation mark next to each highlighted sentence. The second paper icon is labeled distribution, and shows three paragraphs, each with one sentence highlighted. The third paper icon is labeled diversity, and shows four sentences, highlighted in different colors.}
        \label{fig:highlight_guidelines}
    \end{figure}


\emph{D5. Supply enough highlights.}
In our preliminary usability studies, readers often felt uncomfortable when they saw long, unhighlighted passages where they thought important information likely could be found. Some readers wanted to see highlights distributed more uniformly throughout the paper (as opposed to highlights concentrated primarily in an introduction or conclusion). We suggest the rule of thumb that a highlight should be provided around once per paragraph, and that readers should be able to request additional highlights in particularly dense passages.

\emph{D6. Accuracy is crucial.}
A side effect of introducing faceted highlights (where highlights are color-coded by their predicted rhetorical category) was that classification errors became obvious to readers, such as when  a passage about results was labeled as being about methods. Readers found themselves distracted when the classification of a passage clashed with their expectations and became skeptical of the tool's capabilities (E11, E12). If skimming tools provide faceted highlights, it is especially important to classify these categories correctly.

\emph{D7. Support user control and personalization.}
Readers desired more control over the amount of highlights shown by the prototype. Many suggested that the tool could help them fine-tune what was highlighted, either through manual adjustments, or with adaptive personalization of the highlights (i.e., responding to passages a reader has highlighted themselves or highlights they have deleted) (E5, E7, E8, E12). 

A final takeaway from our formative research was that readers believed their comfort using intelligent highlights would change over time, as they became more familiar with the features, the colors associated with the highlights, and the accuracy of the highlights. One reader described this as the issue of ``getting used to seeing highlights that aren't my own'' (E13). This observation motivated our choice of a longitudinal diary study as one of the summative evaluation methods for \scim{} (see Section~\ref{sec:diary_study}).
\section{\scim{}} \label{sec:scim_ui}


\begin{figure*}[t]
        \centering
        \includegraphics[width=0.95\textwidth]{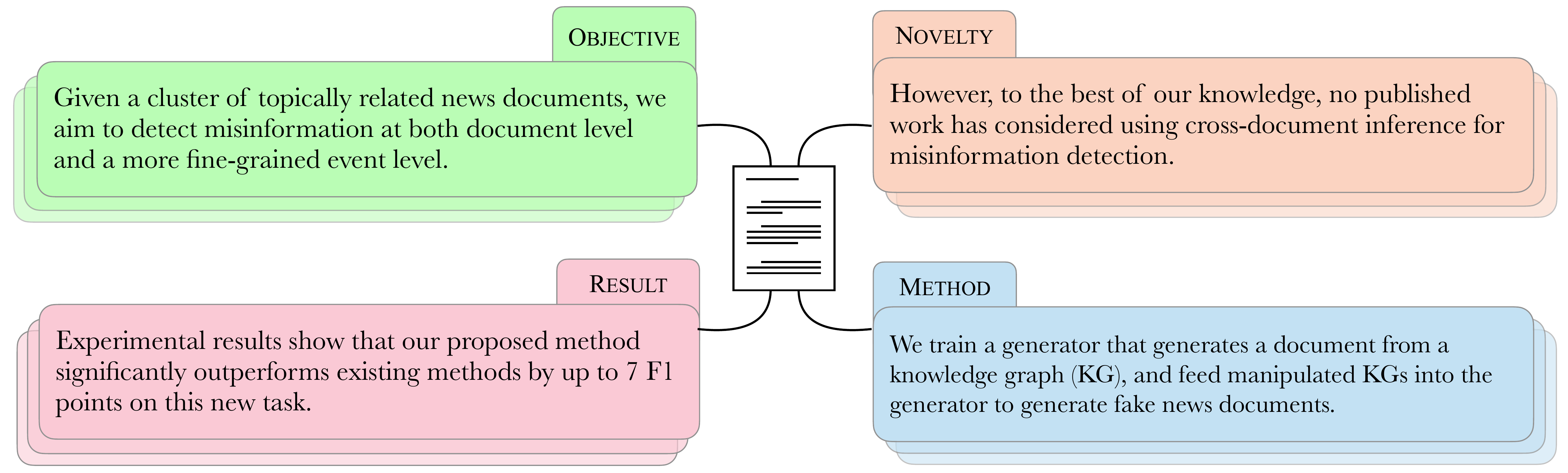}
        \caption{\scim{} classifies and highlights four facets of information commonly found in papers: \textsc{Objective}, \textsc{Novelty}, \textsc{Method}, and \textsc{Result}. These facets aim to surface specific kinds of paper content that align with common skimming goals identified in formative research, reflecting design guideline \textit{D2}. Above, we show example passages matching each of the four facets. The passages appear in Wu et al.'s scientific paper, ``Cross-document Misinformation Detection based on Event Graph Reasoning''~\cite{wu-etal-2022-cross}.}
        \Description{Four colored boxes are connected to an icon of a paper in the center. The four boxes provide an example sentence for each of four facets of information: objective in a green box in the top left, novelty in an orange box in the top right, result in a red box in the bottom left, and method in a blue box in the bottom right.}
        \label{fig:faceted_highlights}
    \end{figure*}

We now describe the design of \scim{}, an interface that provides intelligent support for skimming scientific papers, and explain how particular aspects of the system address the design goals (D1--7) introduced previously.

\subsection{Overview}
A reader interacts with \scim{} as a tool that supports and augments their typical skimming process (D1). One common strategy for readers is to begin with a paper's title and abstract, followed by a piecemeal review of the paper. A reader using this strategy may at the same time follow the highlights offered by \scim{}, which extend into parts of the paragraph a reader may not notice otherwise (D3).

\subsection{Faceted Highlights}
\scim{} intelligently highlights a paper to direct a reader's attention to key passages (Figure~\ref{fig:user_interface}.A). These highlights were tailored in three ways to support skimming.

\textbf{Faceted}. Because readers have different goals when skimming, \scim{} colorizes highlights according to facets of information (D2). To promote memorability we limit the number of facets to four. The specific set of facets was selected to encompass the kinds of information participants described in the formative study, balanced by the requirement that we could detect them reliably (D6), as described in the implementation section.

Numerous schemes exist for sentence-level classification of scientific literature into facets. Coarse-grained schemes classify sentences according to common section names from scientific papers (e.g.,~\cite{hirohata_identifying_2008, cohan_pretrained_2019}) and consist of a small number of facets. Other fine-grained schemes rely on argumentative zones and conceptual structure (e.g.,~\cite{teufel_articles_2002, teufel_towards_2009, liakata_corpora_2010, liakata_automatic_2012}).

We derived a taxonomy of four facets by augmenting facets from of one coarse-grained schema for classifying scientific abstracts~\cite{cohan_pretrained_2019} with the ``\texttt{NOV\_ADV}'' category (i.e., corresponding to sentences describing the novelty of a paper) from Argumentative Zoning~\cite{teufel_towards_2009}. As shown in Figure~\ref{fig:faceted_highlights}, \scim{}'s four facets are: {\colorbox{objectivegreen}{\textsc{Objective}}} (green), {\colorbox{noveltyorange}{\textsc{Novelty}}} (orange), {\colorbox{methodblue}{\textsc{Method}}} (blue), and {\colorbox{resultred}{\textsc{Result}}} (red), each of which is represented in \scim{} with its own color.

\textbf{Low distraction}. Text is highlighted using the familiar paradigm of a solid rectangular box behind the text, since this was observed in our evaluations of prototypes to be noticeable yet minimally distracting (D4). By using the same facet color mapping across papers, we hoped to foster a learned association device for each facet (D4). To help readers develop familiarity with highlight colors, \scim{}'s interface header includes a legend mapping colors to facets.

\textbf{Distributed}. Since users of initial prototypes were concerned when they saw passages without any highlights (D5), we post-processed model predictions to distribute highlights approximately evenly throughout the paper.

\subsection{Controls}
Different readers may have very different goals in skimming, and even a single reader's goals may vary from one passage to the next. To provide flexibility in the skimming experience (D7), \scim{} provides two kinds of controls:

\textbf{Paper-level controls}. If a reader wishes to perform a cursory high-level skim of a paper, they can reduce the density of highlights, or to inspect a paper more closely, they can increase the density. If a reader does not wish to review a particular kind of content as they skim (e.g., they want to learn about the results of a study but not its methodology), they can disable highlights of a certain facet. Readers can control the density of highlights using facet-specific sliders found in \scim{}'s side bar (Figure~\ref{fig:user_interface}.C). As a reader drags a slider, they can see the effect on highlight density as highlights appear and disappear in the paper, markers appear and disappear in the scrollbar, and a count of highlights change next to the slider.

\textbf{Paragraph-level controls}. If a reader desires additional highlights (e.g., if they have encountered a long paragraph of results they wish to skim more closely), \scim{} provides paragraph-level controls allowing them to rapidly access additional highlights. Readers can request more or fewer highlights by hovering their mouse over a block of text, and then clicking on ``+'' and ``-'' buttons that appear in the margin (Figure~\ref{fig:user_interface}.B). This feature provides quick and flexible control to complement paper-level controls, allowing a reader to request highlights precisely where they need them. For both paper- and paragraph-level controls, highlights are added and removed using a sentence prioritization score assigned during the document processing phase, as described in Section~\ref{sec:implementation}.

\subsection{Scrollbar Annotations} \label{sec:scrollbar}
A reader can discover where to skim in a paper by viewing highlight annotations in the scrollbar (Figure~\ref{fig:user_interface}.D). This feature is inspired by edit wear and read wear affordances~\cite{hill_edit_1992} and scrollbar annotations in code editors (e.g., ~\cite{microsoft_scroll_2022}). When viewed together, these annotations can suggest paper structure, for instance implying if a paper has a particularly lengthy methods or results section, and where to find that information. The annotations also offer feedback to readers as they configure highlight density with \scim{}'s controls.

\subsection{Side Bar Display of Faceted Highlights}
A reader may also review a paper's key passages by opening a side bar, which shows a compact list of all highlighted passages in order, grouped by paper section (Figure~\ref{fig:user_interface}.E). This display updates dynamically as a reader configures the highlight density. A vertical colored bar appears next to each passage, providing a subtle and compact indication of the passage's facet. A reader desiring more context for a passage can click on it to scroll \scim{} to the passage's position in the paper, an interaction we call \textit{context linking}.
\section{Implementation} \label{sec:implementation}


\begin{figure*}[t]
        \centering
        \includegraphics[width=0.95\textwidth]{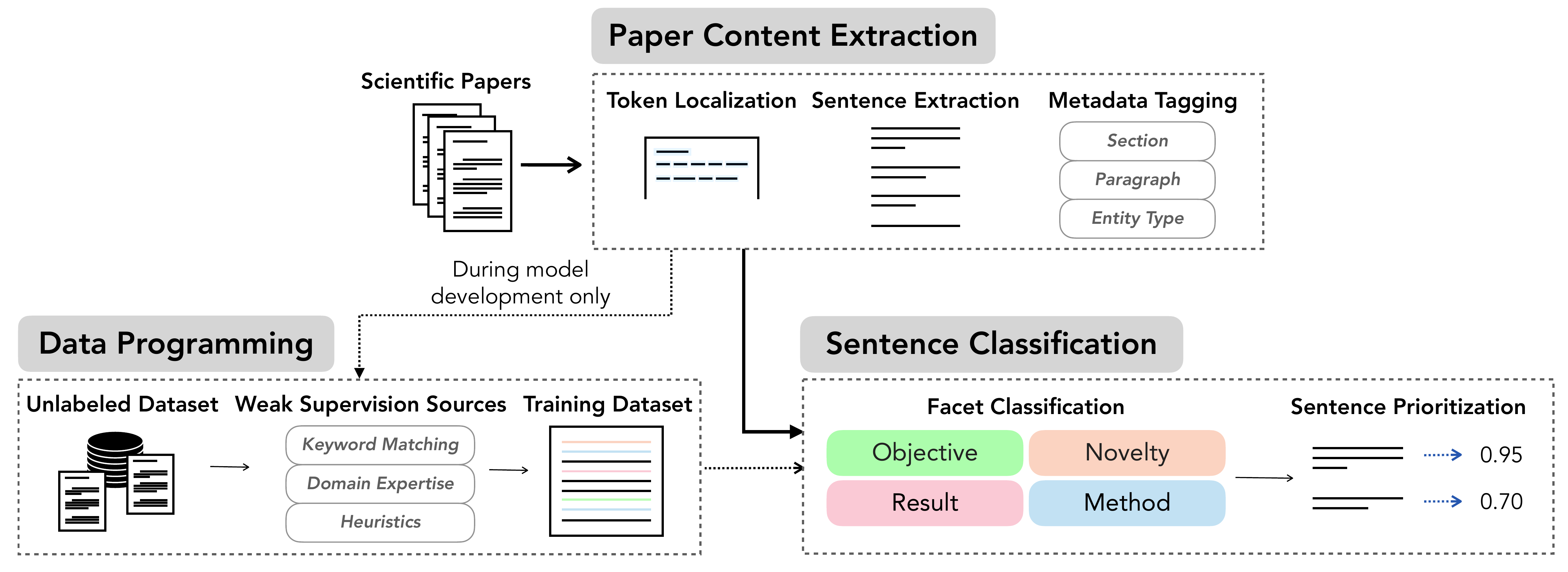}
        \caption{Overview of \scim{}'s paper processing pipeline. \scim{} takes as input a scientific paper in PDF format and then parses it into sentences with bounding boxes and other accompanying metadata. It then classifies sentences into one of four facets using a large language model fine-tuned via a data programming approach. \scim{} chooses which highlights to show by reconciling prediction weights with heuristics controlling highlight distribution and readers' preferences.}
        \Description{System diagram for Scim, containing three major modules: paper component extraction, data programming, and sentence classification. Scientific papers are fed into paper component extraction, which performs token location, sentence extraction, and metadata tagging. During model development only, output from paper component extraction is fed into data programming, the module which uses weak supervision sources to create a weakly labeled training dataset. Output from data programming and paper component extraction are fed into sentence classification, which includes facet classification and sentence prioritization.}
        \label{fig:system_diagram}
    \end{figure*}

\scim{} was developed with an end-to-end document processing pipeline that supports intelligent highlighting. The main component of this pipeline is a pretrained language model, fine-tuned via weak supervision to identify and classify salient sentences within papers. An overview of the pipeline is presented in Figure~\ref{fig:system_diagram}.

\subsection{Paper Content Extraction}
Given an input PDF document, \scim{} uses the open-source Multimodal Document Analysis (MMDA) library~\cite{mmda_2022} to extract textual tokens, mathematical symbols, section headers, and metadata. \scim{} then segments the tokens into sentences, simultaneously merging bounding boxes for tokens into bounding boxes for sentences. Each sentence is labeled with its corresponding section header and paragraph index, attributes which are later used in the prioritization of sentences for which highlights should be shown.

\subsection{Sentence Classification}
To classify sentences into facets, we adapted the sequential sentence classification model from~\citet{cohan_pretrained_2019}, replacing the base BERT model with a pretrained MiniLM model~\cite{wang_minilm_2020,wang-etal-2021-minilmv2}. The MiniLM model considers surrounding context---up to a combined sequence length of 512 words or 10 sentences---when classifying a target sentence. We fine-tuned the model with the \textsc{CSAbstruct} dataset~\cite{cohan_pretrained_2019}, a corpus of abstracts from computer science papers with manually-curated ``gold'' labels. Since sentences only came from paper abstracts, we ultimately found the model insufficient for classifying sentences from the body of papers, so we pursued additional fine-tuning as we describe in the next section.

\subsubsection{Data Programming}
We initially attempted to create manually-curated datasets of ``gold'' facet labels for sentences from full papers. However, this task was difficult to define, time-consuming, and expensive to execute during our pilot runs of the data collection process. As a result, we decided to extend our dataset with weak supervision following a data programming approach~\cite{ratner_dataProgramming_2016} to further fine-tune the model. Weak supervision provides a model-agnostic way to incorporate domain expertise into a model, and is sometimes a satisficing alternative to costly manual annotation. Weak supervision assumes access to a large unlabeled dataset and one or more labeling functions (e.g., heuristics encapsulating domain expertise, crowdsourcing, or knowledge bases), which are used to generate noisy labels for the dataset. While a collection of labeling functions can on their own serve as a classifier, we sought generalization beyond precise but potentially brittle labeling rules. We therefore employed a data programming paradigm to unify and de-noise the labeling functions, creating a weakly-labeled training set of sentences for downstream fine-tuning.

To build an unlabeled dataset for weak supervision, we extracted full paper sentences from the proceedings of NAACL 2018, 2019, and 2021, and ACL 2020--2022. In total, the dataset consisted of 3,051 papers with 606,400 unlabeled sentences. We then created weak supervision labeling functions consisting of heuristic rules and keyword matches to provide noisy facet labels for sentences in the dataset. For example, one rule-based supervision function detected sentence salience based on the presence of author intent via keywords such as ``we'', ``our'', ``this paper,'' and their aliases. Other labeling functions relied on keyword matches to perform facet labeling. For example, sentences were weakly labeled as \textsc{Novelty} if any relevant keywords (e.g., ``novel'', ``propose'', ``differ,'' and their aliases) could be found. We used Snorkel~\cite{ratner_snorkel_2017} to unify these labeling functions and output a dataset of weakly labeled sentences.

The dataset was further improved by incorporating weakly labeled negative sentences, selected from the full papers associated with the \textsc{CSAbstruct} abstracts used during the first round of training. We selected novel sentences by using the \texttt{all-mpnet-base-v2} model~\cite{song2020mpnet} from the Sentence Transformers library~\cite{reimers-2019-sentence-bert} to score sentence similarity between full text and the abstract, and then labeling the most dissimilar sentences to the abstract, and which were not labeled with a facet in the prior phase, as not relevant for any facet (using an empirically chosen threshold cosine similarity of $0.25$). Model fine-tuning was done on an NVIDIA A6000 GPU, using 0.1 dropout rate and Adam optimizer~\cite{kingma_adam_iclr2015} over 5 epochs, and $5\cdot10^{-5}$ learning rate. All parameters were determined using the \textsc{CSAbstruct} validation split.

\subsubsection{Evaluation}
We conducted a preliminary evaluation of \scim{}'s sentence classification model over a set of 20 NLP papers. We recruited annotators from Upwork, an online freelancing marketplace. All hired annotators were required to have experience with NLP and scientific writing. Detailed instructions asked annotators to role play as a reviewer for a scientific communication magazine, tasked with creating abridged versions of scientific papers. Annotators were asked to identify significant, complete sentences within each of the 20 papers, and were paid \$20 USD/hr.

Each paper took on average 20 minutes to annotate, and was annotated by three Upworkers using the PAWLS PDF annotation tool~\cite{neumann2021pawls}. Sentences selected by at least two of the three annotators were considered as ground truth ``significant sentences,'' and collected into a test set. On this test set, our classification model achieved an F1 score of 0.533, compared to an annotator-annotator F1 score of 0.725 (which we consider as a gold-standard, i.e., a performance ceiling, since there is inherent variability in which sentences annotators believe are significant for skimming). Our goal with this preliminary evaluation was not to evaluate whether we advanced the state-of-the-art in NLP, but rather to verify that the model reliably identified meaningful highlights for use in \scim{}.

\subsection{Cleaning and Prioritizing Highlights}
\scim's user interface selects which highlights to show using the predicted facet label, probability score, and other heuristics. One heuristic enforced consistency between facet labels and the section in which a sentence appeared (e.g., if a highlight appeared within a methods section, it had to be tagged with the ``\textsc{Method}'' facet; similar constraints were imposed for the ``\textsc{Novelty}'' and ``\textsc{Results}'' facets). Another heuristic prompted a more uniform distribution of highlights throughout a paper, prioritizing sentences within paragraphs which did not already contain other highlights.

\subsection{User Interface Implementation}
\scim{} is implemented as a web application built atop the PDF rendering platform \verb|pdf.js|~\cite{mozilla_pdfjs_2022}. The system retains text markup already present in the paper which may support skimming, such as hyperlinks, clickable citations, bold and italicized text, and other visual cues provided by the authors. \scim{}'s features including highlights, side bars, and controls were implemented as interactive React components incorporating widgets from the Material UI library~\cite{materialui}.
\section{Study 1: \textmd{In-Lab Usability Study}} \label{sec:lab_study}

We first conducted an in-lab usability study to assess how \scim{} affects readers' ability to search for specific kinds of information in a paper. Participants in the study were asked to complete a series of short tasks using both \scim{} and a standard document reader. Our usability study sought to answer two research questions:

\indent \textbf{RQ1.} \textit{Does \scim{} enable readers to skim papers more quickly?} \\
\indent \textbf{RQ2.} \textit{How does \scim{} affect readers' ability to identify relevant information after a skim?}

\subsection{Study Design}

\subsubsection{Participants}
We recruited 19 participants (8 male, 10 female, 1 non-binary) via university-affiliated mailing lists and Slack channels. We also conducted pilot studies with three additional participants, results of which we do not include in our analysis. Participants were required to have experience reading NLP papers, since they would be required to do so during the study. They ranged from 21 to 30 years of age, and included 11 PhD students, 5 master's students, 2 software engineers, and 1 industry researcher. Participants self-reported an average of 3.78 (on a 5-point Likert scale) for comfort with reading NLP papers, suggesting they were generally familiar with the type of literature used in the study. Participants were compensated $\$25$ USD for their time.

\subsubsection{Procedure}
Participants first provided consent and then were led through a tutorial of \scim{}'s features. The study used a within-subjects design, and consisted of three tasks, each with two sub-tasks, one for each of the two reading interface conditions---\scim{} and a standard document reader. We designed the study to be completed in under one hour to limit participant fatigue. The studies were conducted remotely via Zoom. To minimize biases, we counterbalanced the order of the reading interfaces and papers used in each task. Below, we describe the three tasks.
\begin{itemize}
    \item \textit{Task 1}: Participants skimmed a paper and identified a passage in the paper that described a key feature (e.g., dataset creation or evaluation) of the paper. This task was intended to familiarize participants with the two interfaces, so we did not include any measures from this task in our analysis.
    \item \textit{Task 2}: Participants skimmed a paper and answered two multiple-choice questions based on information found in the paper. Answers to these questions \textit{could} be found in text highlighted by \scim{}. We hypothesized the main points highlighted by \scim{} should be easier to locate, and this task was designed to test that hypothesis.
    \item \textit{Task 3}: Participants skimmed a paper and answered two multiple-choice questions based on information found in the paper. Answers to these questions \textit{could not} be found in text highlighted by \scim{}. In contrast to Task 2, we hypothesized that \scim{} might prove a hindrance when finding information outside of the highlights, and this task was designed to check this concern.
\end{itemize}

Participants skimmed a different paper for each of the sub-tasks. The six papers for these tasks~\cite{abrams_social_2022, li_multispanqa_2022, stasaski_semantic-diversity_2022, sulem_yes-no-idk_2022, ibraheem_putting_2022, xie_word_2022} were selected from the proceedings of NAACL 2022, and corresponded to the following types: (1) technical papers introducing new datasets or metrics, (2) exploratory papers investigating the effectiveness of current tools and proposing new design guidelines, and (3) technical papers proposing novel language models for specific applications. Questions in Tasks 2 and 3 focused on aspects of a paper a reader might be interested in while skimming, such as evaluation metrics or the motivation behind a proposed method. Regardless of interface, participants were given multiple attempts and asked to skim until they answered correctly.

For each question, we used the following quantitative metrics:
\begin{itemize}
\item \textit{Time ---} {The number of seconds taken by the participant to answer the question, from when the paper was first opened to when the correct answer choice was selected.}
\item \textit{Accuracy ---} {A binary variable indicating whether the participant's first response to the question was correct.}
\item \textit{Difficulty ---} {A five-point Likert scale variable indicating the participant's self-assessment of the following prompt: ``I found the task difficult.''}
\end{itemize}

After completing all sub-tasks, participants were also asked to self-assess on a five-point Likert scale whether they found the tasks overall easier to complete with \scim{} and whether \scim{}'s highlights were distracting during skimming.


\begin{figure*}[t]
        \centering
        \begin{subfigure}{0.55\textwidth}
            \centering
            \includegraphics[width=\textwidth]{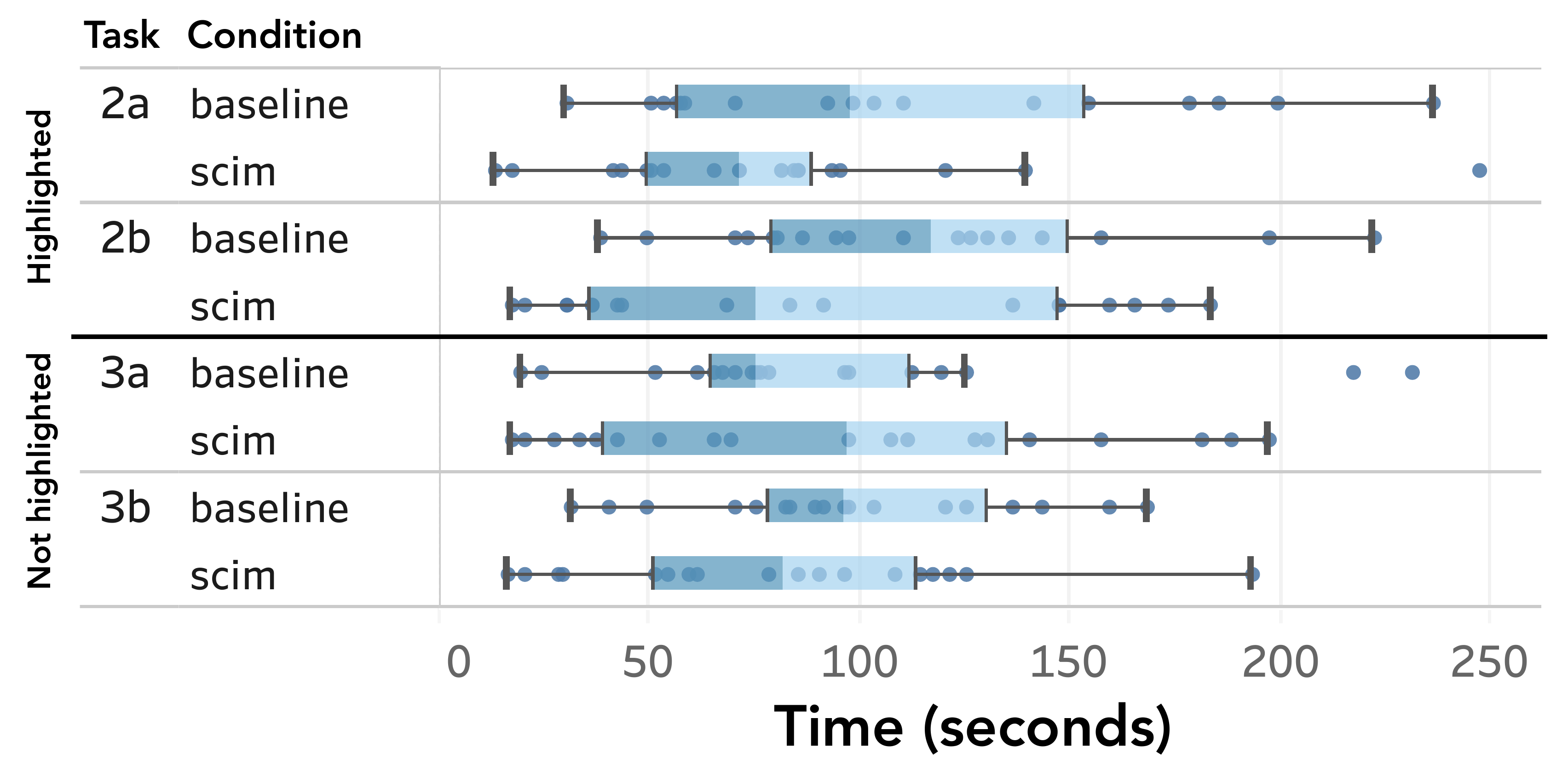}
            \label{fig:lab_study_timing}
        \end{subfigure}
         \hfill
        \begin{subfigure}{0.44\textwidth}
            \centering
            \includegraphics[width=\textwidth]{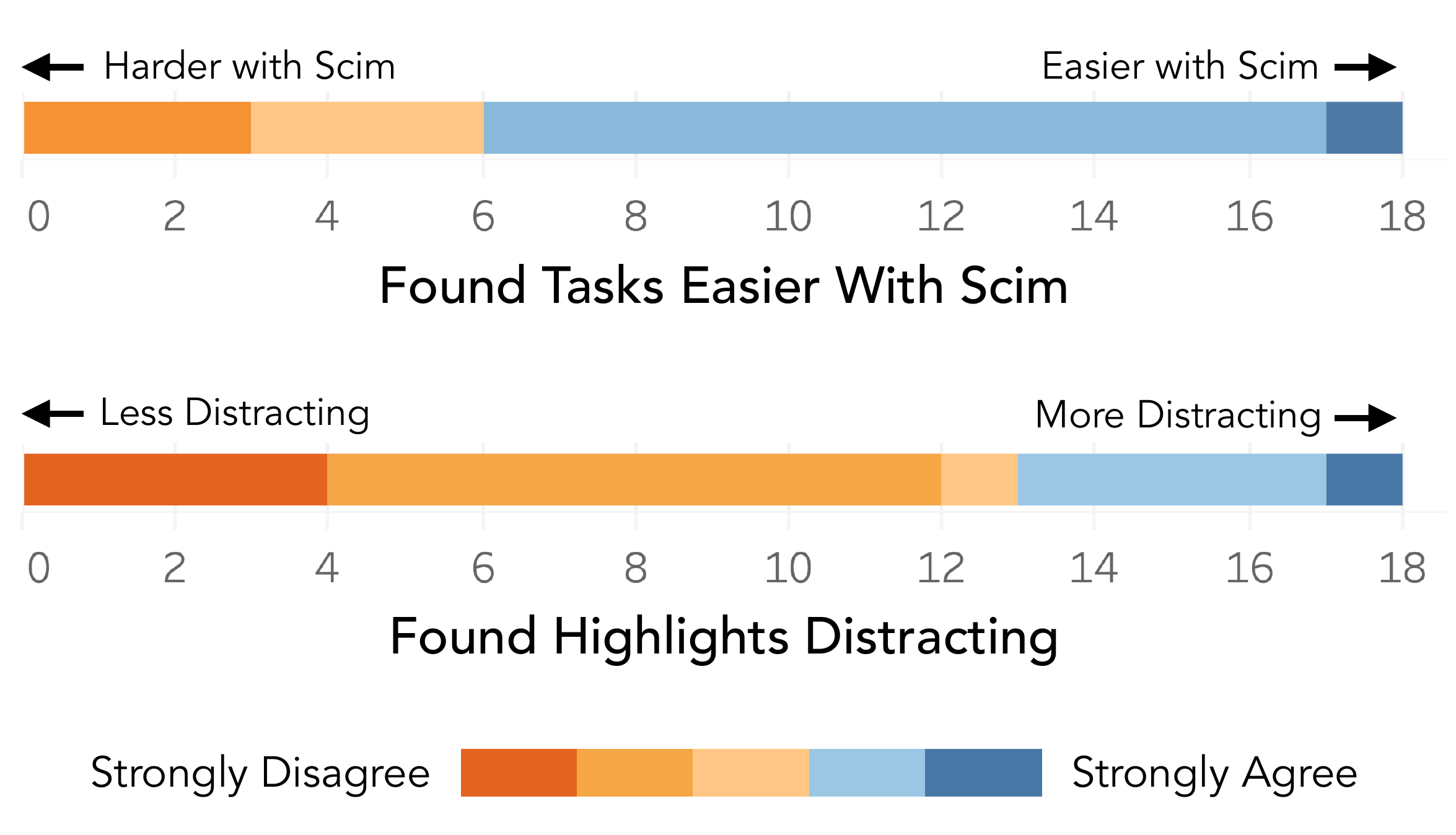}
            \label{fig:lab_study_likert}
        \end{subfigure}
        \caption{
            \textbf{(Left)} Time taken by participants to complete each information seeking question in Tasks 2 and 3 of the in-lab usability study. Overall, participants answered questions more quickly with \scim{} than with a standard (baseline) document reader.
            \textbf{(Right)} Participants' subjective responses regarding the ease of completing information seeking tasks with \scim{} compared to a standard document reader, and whether they found \scim{}'s highlights distracting.
        }
        \Description{On the left, a box and whisker plot shows participant response times to task questions for the two conditions, baseline and Scim. Median times are lower for Scim in Tasks 2a, 2b, and 3b, and lower for baseline in Task 3a. Overall, times for Scim are lower and distributed further left than the baseline. On the right, a horizontal stacked bar chart shows aggregated participant self-assessments to task difficulty and highlight distraction, on a five point Likert scale. Over half of the responses agree that tasks were easier to complete with Scim. Over half of the responses did not find highlights distracting.}
        \label{fig:lab_study_results}
    \end{figure*}

\subsubsection{Analysis}
We compared readers' time, accuracy, and perceived task difficulty using linear mixed-effects models~\cite{lme_lindstrom_1990} with reading interface as a fixed effect, task and question number as nested fixed effects, and participant as a random effect. We first conducted \textit{F}-tests for any differences across the interface conditions, and then we conducted post-hoc \textit{t}-tests when appropriate for differences in the estimated fixed-effects between conditions.

\subsection{Results}
Participants answered questions more quickly with \scim{} ($\mu$~=~94.3s, $\sigma$~=~74.9s) than with a standard document reader ($\mu$~=~117.7s, $\sigma$~=~76.4s). This difference was significant (F(1, 126)~=~4.17, p < .05). The difference was more pronounced in Task~2, where the correct answer was located within one of the highlights (F(1, 54)~=~4.84, p < .05): readers took an average of 93.8s with \scim{} ($\sigma$~=~81.6s) versus 127.3s with the standard reader ($\sigma$~=~77.8s). In Task~3, where the correct answer was not located within one of the highlights, there was no significant difference in the time taken (F(1, 54)~=~0.58, p~=~.45), with participants taking 94.8s with \scim{} ($\sigma$~=~68.7s) versus 108.0s with the standard reader ($\sigma$~=~74.7s).

There was no significant difference \mbox{(F(1, 126)~=~0.22, p~=~.64)} in participants' accuracy with \scim{} ($\mu$~=~0.80, $\sigma$~=~0.40) compared to a standard document reader ($\mu$~=~0.76, $\sigma$~=~0.43). There was also no significant difference (F(1, 119)~=~0.01, p~=~.92) in readers' perceived difficulty in answering questions with \scim{} \mbox{($\mu$~=~2.32, $\sigma$~=~0.89,} with 5.00 indicating strong difficulty) compared to a standard document reader ($\mu$~=~2.31, $\sigma$~=~1.02). Altogether, the results show that \scim{} reduced the time it took for readers to seek information in papers, with no observed significant difference in accuracy or effort.
\section{Study 2: \textmd{Longitudinal Diary Study}} \label{sec:diary_study}

Participants in our lab usability studies noted that it would take some time to acclimate to a novel reading interface like \scim{} before they felt comfortable using it. To better understand realistic long-term use, we therefore also conducted a two-week long diary study. This study let readers use \scim{} for papers of their choice from a list relevant to their discipline, leading to alignment of their motivation with typical motivations for skimming. Participants could choose when they read, and for how long, as long as they skimmed using \scim{} at least once a day. We designed the diary study to provide insight into the following research questions:

\begin{enumerate}
    \item[\textbf{RQ1.}] \textit{What value can intelligent highlight-based skimming aids provide to researchers?}
    \item[\textbf{RQ2.}] \textit{How do researchers make use of skimming aids as they read?}
    \item[\textbf{RQ3.}] \textit{In what scenarios do researchers find skimming aids useful?}
    \item[\textbf{RQ4.}] \textit{What are the limitations of highlight-based skimming aids?}
    \item[\textbf{RQ5.}] \textit{What features should future intelligent skimming tools have?}
\end{enumerate}

\subsection{Study Design}

\subsubsection{Participants}
We recruited participants through university-affiliated mailing lists, Slack channels, and public posts from the authors' Twitter accounts. Participants were required to have prior experience reading or writing research papers. Preference was given to those with experience reading papers in the field of natural language processing (NLP), because the collection of papers we preprocessed for this study came from a recent NLP conference. A total of 12 participants were recruited for the study (6 male, 6 female). Two were master's students, and ten were PhD students. PhD students spanned a range of experience, with 1 first-year student, 3 second-year students, 2 third-year students, 3 fourth-year students, and 1 fifth-year student. No participants had participated in any of the prior lab studies. Participants were compensated $\$100$ USD at the end of the study.

\subsubsection{Reading Materials}
Though \scim{}'s pipeline was able to process arbitrary papers within a few seconds, we wished to reduce the time it took for participants to load papers during the diary study. As a result, we preprocessed a set of papers we felt would be exciting for participants to read, specifically the proceedings of NAACL 2022, one of the most recent and widely-read NLP conferences. We selected these papers since \scim{} had been fine-tuned primarily on NLP papers, and we expected it would perform appropriately for this collection. We also anticipated the NLP community would provide a broad audience from which we could recruit participants for a diary study. We also allowed participants to request other papers outside of this collection to read with \scim{} throughout the study; in total, an additional 10 NLP papers were preprocessed.

\subsubsection{Procedure}
The diary study consisted of three stages: a welcome session, a two-week long observational period, and an exit interview. During the welcome session, participants completed a tutorial of how to use \scim. They were given a few minutes to try out the interface, and to ask questions. Each participant was also shown the online diary (hosted in a Google Doc), and briefed on the protocol for recording their skimming experiences.

Then, during the observational period, participants were asked to spend 5--10 minutes each day, for 10 days, skimming at least one paper and completing a structured reflection in the diary.\footnote{Nearly all participants succeeded in completing 10 days of diary entries. Only 1 of 12 failed to complete all required entries; they completed only 7 of 10. When a participant fell behind in their diary entries, we sent them light email reminders.} On the first day of the observational period, participants skimmed papers using a standard (non-\scim{}) document reader for the first day to provide a point of comparison with \scim{}. During the subsequent nine days, they skimmed papers with \scim{}. Following each skimming session, they completed a diary entry, consisting of the following questions:
\begin{enumerate}
    \item Which papers did you skim today, and how long did you spend skimming each one?
    \item What highlights (if any) drew your attention to something you might have missed without the highlights?
    \item Did highlights help you skim this paper? Explain.
    \item List one or more ways the system could have helped you better skim this paper.
\end{enumerate}

After the observational period, we conducted exit interviews with participants. They were asked to reflect on their experience using \scim{} in detail, including how it supported their skimming and opportunities for improvement.

\subsubsection{Analysis}
We conducted a thematic analysis on the qualitative data---diary entries and transcripts from exit interviews---following the approach described by~\citet{creswell_qualitative_2016}. One author identified significant excerpts from the diary entries and transcripts, and iteratively developed and refined a set of themes represented in the data. A second author validated the analysis by reviewing the themes, checking their alignment with the excerpts, and proposing revisions. A total of 177 responses to diary prompts were analyzed (participants left responses to some questions blank). We also instrumented and analyzed behavioral logs detailing interactions with \scim{} for each participant. In reporting results, we refer to participants with the pseudonyms P1--P12. The utterances presented below were edited to elide identifying information while preserving their meaning.

\subsection{Results}

In this section, we present the findings of our diary study, organized by their relevance to each of our five research questions.


\begin{table*}[t]
    \caption{A summary of usage of \scim{}'s features during the diary study. Notably, most readers used most features at least once. Use of the highlight controls varied widely, with some readers using them heavily (P6, P10, P12), and others less often (P1--3, P6, P7--9). All readers used the highlight browser on multiple occasions. Faceted highlights are omitted from this table, because we could not collect log data as to when readers looked at highlights.}
    \label{tab:feature_usage}
    \begin{tabular}{r | c c c c c c c c c c c c}
        \toprule
        \textbf{Feature} & \textbf{P1} & \textbf{P2} & \textbf{P3} & \textbf{P4} & \textbf{P5} & \textbf{P6} & \textbf{P7} & \textbf{P8} & \textbf{P9} & \textbf{P10} & \textbf{P11} & \textbf{P12} \\
        \midrule
        Highlight Browser           &5 & 10  & 8   & 10   & 9   & 4  & 12   & 3   & 5  & 20   & 6  & 19 \\
        Global Highlight Controls   & 1  & 0 & 3    & 0   & 2  & 16   & 3  & 3   & 0  & 22  & 9   & 4 \\
        Local Highlight Controls    & 3  & 2  & 1    & 8   & 0  & 12   & 0   & 0   & 0  & 16   & 3  & 34 \\
        Context Linking             & 0  & 1  & 0   & 3  & 30   & 2   & 0   & 0   & 0  & 6   & 0  & 8 \\
        \bottomrule
    \end{tabular}
\end{table*}

\subsubsection{The value of \scim{} as a skimming aid (RQ1)}\label{sec:value} \hfill

For many readers, \scim{} helped with skimming by allowing them to focus their attention and attain a high-level understanding of the paper (P5, P6, P9, P10). Furthermore, \scim{} helped readers identify key concepts and review the main ideas of papers. P5 described \scim{} as guiding her to the important contributions of the papers she skimmed, and the highlights as offering a ``gist of the paper beyond what was in the abstract.''

And though highlights helped readers review the paper as a whole, they could also help them orient to specific aspects of a paper they wanted to understand. For instance, P1 and P2 both noted that the highlights helped them to understand the results of the paper more quickly, which are often quite dense and text-heavy. \scim{}'s highlights also helped readers attend to interesting details in sections of papers they might have otherwise skipped over (P1, P4, P11). This was described as ``slowing down'' and skimming with greater care:
\begin{quote}
    \textit{This was a paper that is very light on methods and most content is about results, which I tend to skim over. So the highlights helped me slow down and slightly more carefully read a few places.} (P4)
\end{quote}

For some readers, skimming without \scim{} required two passes, first skimming a paper to identify relevant passages, and then re-reading passages of interest in greater detail (P5, P8). \scim{} could alleviate the need for multiple passes:
\begin{quote}
    \textit{With highlights, I usually spend more time reading and understanding the highlighted content and skimming the other content. Without the highlight[s], I need to scan the entire content first, identify the critical points and then understand it. The highlights save me time in skimming the whole paper.} (P8)
\end{quote}

\subsubsection{How researchers made use of \scim{} (RQ2)} \hfill

Usage of \scim{} entailed usage of its constituent features of highlights, the highlight browser, controls, and context linking. All readers made use of most features at least once (Table~\ref{tab:feature_usage}). We surmise that the most frequently used feature was the highlights, for several reasons. First, the feature was always turned on. Second, highlights figured prominently in our conversations with readers, as evidenced by the rest of this section. And third, most readers reported that the highlights helped them find useful information during their daily readings (see Section~\ref{sec:useful}).

While the predominant method of interaction with \scim{} was likely to view highlights within the paper, a second commonly used feature was the highlight browser: all readers opened \scim{}'s side bar more than once, with the average reader opening it 9.3 times. Readers described the highlight browser as supporting navigation and providing a rapid understanding of paper contents (P7, P9, P10). It was also described as an ``extractive summary'' (P2). One reader thought the highlight browser provided a ``better way to skim'' in comparison to highlights, which at the time of their diary entry, they believed made the paper ``difficult to read'' (P7).

Nearly all readers used both global and local controls to configure the number of highlights. Global controls were typically used a small handful of times to achieve an acceptable density of highlights (which was then persisted into subsequent skimming sessions). Only a few readers adjusted the highlights via the global controls across multiple papers skimmed. When asked, readers typically reported that the default density of highlights was appropriate (P2, P6, P7). That said, most did adjust the number of highlights with paper-level controls at least once. Exit interviews confirmed that readers tended to tune the level of highlights to the preferred level on the first day of use. One participant asked for highlight controls with coarser options, for instance enabling them to toggle between one mode showing only the most important highlights, and another with many highlights for a deeper skim.

Readers seemed to use \scim{} to augment, rather than replace, their existing skimming strategies. Readers reported directing their attention both to the highlights and to conventional paper landmarks like section headers and visual content. For example, P9 described their process as navigating through the main sections of a paper as they might in a typical skim, and then using the highlights to identify important information within those sections. P4 similarly described skimming using the combination of section headers and highlights.

For some readers, it took some time to become accustomed to using \scim{} (P8, P10). One issue seemed to be developing trust in what was highlighted (P8). In their exit interview, P10 described their how their trust and interactions with \scim{} evolved over the course of the study:
\begin{quote}
    \textit{I feel like I just got more used to the highlights. ... When I would see an objective highlight, I would trust it. I found the results highlights to be very helpful, so I would immediately focus on those. I would open the side panel right away instead of waiting during the end of the paper. I just got used to the tool, and I learned how to use it fast, depending on the paper and what I wanted to get from the paper.} (P10)
\end{quote}

\subsubsection{Circumstances in which \scim{} was useful (RQ3)}\label{sec:useful} \hfill

Overall, the intelligent highlights appeared to be useful during a majority of skimming sessions. In response to the diary question, ``Did highlights help you skim this paper?'' 74 of 105 (70.4\%) responses answered in the affirmative. There were a handful of circumstances in which readers reported \scim{} as particularly useful.

One circumstance where \scim{} was useful was in reading dense passages of text. The highlights made long passages that were absent of ``visual support'' such as figures more approachable (P3, P5). \scim{} helped one reader skim a detailed experimental section and identify several important details which, due to the density of text, they ``might have skipped if not for the highlights'' (P5). Readers reported \scim{} as helpful not just for dense passages, but also for papers that were text-heavy as a whole, such as survey papers (P5, P11).

Intelligent highlights were seen as useful to readers who sought information from papers on a topic they did not typically read about (P8, P10), assisting them in identifying and focusing on important paper content:
\begin{quote}
    \textit{For me it was also generally useful for reading papers that were a little out of my comfort zone. \ldots{} In that case the highlighting helped me focus on, read, and conceptualize better certain parts of the methodology in order to better understand the conclusions.} (P10)
\end{quote}

The highlights also provided a summary of the paper in their own right. One reader described a situation where they were ``not particularly interested in this paper.'' For them, the highlights served as ``a summary'' that they could read in lieu of looking closely at the paper (P4). This suggests an interesting possibility for intelligent highlights to help not just highly-motivated skimmers, but also those skimming papers in lower-motivation contexts.

\subsubsection{Limitations of \scim{}'s model of intelligent highlighting (RQ4)} \hfill

Readers identified several ways that intelligent highlights might be extended to be made more useful. One concern was that highlighted passages sometimes lacked sufficient context to be understood alone (P1--3, P7--8, P11).

\begin{quote}
    \textit{When reading the highlights, the context is often missing. Sometimes it is just in the lines before and after, but sometimes we need to find it which then makes reading difficult as there is now more back and forth instead of a linear reading.} (P7)
\end{quote}

\scim{} was designed with the hope that readers would look for such ``context'' in the surrounding text by simply moving their focus from highlighted to unhighlighted text. In practice, it could be disruptive for readers to seek out this context. Necessary context could appear just before or after the highlight in the paragraph, and in some cases even in other sections. For one reader, skimming highlights that lacked context therefore became a process that resembled ``more back and forth instead of a linear reading'' (P7).

Some readers desired tighter integration between \scim{}'s highlights and existing visual cues within a paper. While \scim{} did not occlude or hide text that the author had emphasized (e.g., bolded text, section headers, or bulleted lists), this emphasized text was often not highlighted. As a result, readers discovered inconsistencies between the visual cues introduced by authors and the highlights suggested by \scim{} (P2, P4, P5), such as bolded result statements or contributions in a list, which were not consistently highlighted.

Sometimes, text was highlighted in other unexpected and undesired ways. For instance, \scim{} sometimes highlighted only one contribution from a list of bulleted contributions, when readers believed it should have highlighted all of them (P1, P7). \scim{} was also unpredictable when highlighting passages that contained dense math notation (P1, P6, P11), and readers wished for highlights to apply to visual content like tables and figures (P2, P5, P7, P12).

\subsubsection{Envisioning future intelligent skimming tools (RQ5)} \hfill

\scim{} represents just one way in which intelligent assistants could support skimming, and readers described alternative ways that future tools could help them skim. For some readers, \scim{}'s highlights provided too much detail, particularly if they desired only a high-level understanding of the material (P6, P8). Readers suggested that an abstractive summarization of paper content (e.g., ``with a bit of info pulled from tables/graphs/figures/examples'' (P12)), could lessen the effort required to understand dense sections of papers (P1--2, P7--P8, P12). Recent large language models have achieved impressive advances in summarizing scientific texts, and future tools could leverage these models to augment the reading experience with abstractive summaries. Readers also believed they could be aided with better tools for navigation. One reader desired the ability to use a paper's abstract or introduction as an index into related highlights in the rest of the paper (P2). Another reader wished to see the paper summarized in question-and-answer format, realizing they often sought answers to questions while they skimmed, such as ``What are the research questions? What are the novelties/contributions of this study? What data/model/evaluation methods do they use? What are the main results? What are the limitations?'' (P8). While \scim{} addresses these information needs through faceted highlights, future tools could support more conversational interactions between readers and the papers they skim.
\section{Discussion}
In this paper we sought to understand how intelligent user interfaces could support readers in skimming scientific papers. We designed and evaluated \scim{}, a tool that augments the skimming experience with automatic faceted highlighting of paper content. A lab study showed \scim{} reduced the amount of time to complete short information seeking tasks in scientific papers, with no significant difference in readers' self-reported task difficulty. In a subsequent diary study, we observed how researchers might use \scim{} in more realistic settings. Readers believed \scim{} helped them develop a high-level understanding of papers and determine which passages to skim or skip. \scim{} was seen as useful for skimming dense texts and papers from unfamiliar domains. Below, we consider these results, amidst limitations of our research approach, and implications for the agenda of developing intelligent skimming aids.

\subsection{Skimming versus Scanning}
In the lab study, the information seeking tasks we used were intended to measure participants' speed and accuracy while skimming. However, we noticed that participants often exhibited behavior that more closely resembled \textit{scanning}. Unlike skimming which involves a rapid high-level comprehension of text, scanning is a subtly different reading process concerned with locating specific pieces of information within a body of text. In the lab study, some participants utilized conventional scanning strategies such as ``Control+F'', using keywords in the question or answer choices as prompts. While this strategy was typically unsuccessful since the questions were designed with text which nullified this keyword-based scanning strategy, this behavior suggests participants did not necessarily attempt to skim the paper to gain an understanding of the paper to answer the questions, but instead scanned the text for keywords to locate the exact answer to the question. For readers like these, our results may be less indicative that \scim{} helps with the skimming process but rather the scanning process.

\subsection{Supporting Experts and Novices}
\scim{} was designed to help experienced skimmers get more out of skimming. It was not, however, designed to help inexperienced skimmers develop proficiency with skimming. Skimming assistants for teaching skimming may or may not have a lot in common with \scim{}. As AI models become increasingly adept at identifying salient paper content, those models may be useful not just to identify important content, but to coach skimmers to find this content as well. The development of such AIs and accompanying interfaces poses the interesting design problem of ensuring that readers have a consistently productive experience learning the essentials of skimming, regardless of their prior background or the documents they skim.

\subsection{Risks to Attention}
One risk of introducing technologies like \scim{} is that they may have unintended consequences for a reader's attention. Skimming requires significant attention to understand the idiosyncrasies and nuances within papers. Tools that augment the paper with assistive affordances may inadvertently deplete a reader's limited attention if they impose additional cognitive burden, as might be the case if it highlights content in a way readers do not expect. The tools could also lead a reader to pay less attention to the paper as they skim, once they are no longer required to drive the skimming process themselves. Furthermore, should such skimming assistants be readily available, readers may not be incentivized to deeply read papers, but rather enticed to skim by the presence of highlights.

Skimming aids like \scim{} therefore need to be designed in a way that respects a reader's attention and the value of deeply reading. They should be accurate and reliable. They should be deployed alongside studies that understand their effect on readers' engagement with texts. Furthermore, they should be developed and deployed in tandem with tools that support and encourage researchers to deeply read, and in general encourage good reading practices within the research community. For instance, affordances for skimming might be made available only when searching quickly through multiple papers or while reading on the go, but then are limited in scenarios befitting a deeper read.

\subsection{Limitations of Highlights}
Without sophisticated controls and affordances enabling more goal-driven or personalized skimming, highlights only present a single pathway through a paper. Highlighting is a cueing mechanism that directs reader attention and assists in the foraging of information, but it does not address other sensemaking aspects of skimming. As readers suggested in the diary study, there are numerous ways in which skimming aids could provide more holistic support, via additional context for highlighted passages, integration with existing visual cues, highlighting of visual content, complementary usage of abstractive summarization, or enhanced navigation support.

\subsection{Future Work}
We see several opportunities for the research community to further explore the potential of intelligent skimming aids.

\subsubsection{Improving Highlight Quality}
The effectiveness of \scim{}, like that of many other AI-infused user interfaces, is limited by the accuracy of its underlying AI models. Future research could improve the usefulness of highlights with alternative algorithmic approaches. One promising direction may be the use of long-form summarization models sensitive to our highlight-relevant design guidelines, or other large-scale generative models. Features including a paper's hierarchical structure, author-cued content, or visual content might also be leveraged to improve highlighting accuracy. Furthermore, improvements to PDF processing are necessary to improve the user experience for tools like \scim{}. Minor errors in \scim{}'s PDF processing resulted in content like footnotes, section headers, tables, or figures being concatenated with paper sentences, which led both to poor classification of those passages and highlights that included seemingly disparate content. Such issues in PDF processing may need to be resolved to improve classification accuracy and the cleanliness of the highlights' appearance in the reading interface.

\subsubsection{Social Annotations}
Our preliminary studies suggested that readers may be hesitant to adopt an augmented reading interface like \scim{} due to distrust in the AI's ability to provide the relevant highlights. Some readers mentioned that they might trust highlights created by other people (e.g., fellow researchers) more than those generated by AIs. Could social annotations be used to produce better highlights? Social highlights have been extensively explored in other settings, including studies on the effect of social annotation on attention within public multimedia content~\cite{carter_digital-graffiti_2004}, news reading~\cite{kulkarni_newsreading_2013}, and education~\cite{glassman_mudslide_2015, yoon_richreview_2016, zyto_successful_2012}. Modern online publishing platforms such as Medium also show ``popular highlights,'' suggesting the potential for social highlights in reading tools for scientific literature as well. Such affordances might port nicely into a system like \scim{}. The coordination of social highlights with AI-generated highlights could make for an interested area for future research.

\subsubsection{Personalization of Skimming Aids} 
As readers continue to interact with augmented reading interfaces, we envision an opportunity for AI-infused systems to learn from repeated reader interactions, providing personalized and proactive reading support to help mitigate undesirable cognitive overhead introduced by these systems. They could also be tailored to readers' individual reading behaviors by considering their experience reading papers within a particular field, their typical information needs, or their goals for reading a particular paper.

\section{Conclusion}
Our formative research yielded seven goals motivate the design of intelligent tools for skimming scientific papers. We instantiated these motivations in \scim{}, an intelligent skimming interface which supports skimming with faceted, evenly-distributed, minimally intrusive, configurable highlights. A lab usability study found participants located information in papers more quickly with \scim{} than with a standard document reader. In a two-week-long diary study, participants remarked ways in which \scim{} supported a rapid, high-level skimming of papers. \scim{} was found to be particularly useful for dense passages of text and for papers from unfamiliar domains. Altogether, these studies suggest the potential for intelligent tools to support researchers in skimming scientific literature.

\begin{acks}
This project is supported in part by NSF Grant OIA-2033558, NSF RAPID award 2040196, ONR grant N00014-21-1-2707, and the Allen Institute for Artificial Intelligence (AI2).
\end{acks}


\balance
\bibliographystyle{ACM-Reference-Format}
\bibliography{main}

\end{document}